\documentstyle[preprint,aps,epsf]{revtex}
\tighten
\preprint{\parbox[t]{40mm}{SHEP-96-32\\
          hep-ph/9611390}}
\begin{document}
\draft
\title{Chiral Symmetry Breaking in Axial Gauge QCD from the
Dyson-Schwinger Equations}
\author{A. J. Gentles and D. A. Ross}
%\vspace*{0.3\baselineskip}
\address{Department of Physics, University of Southampton, Highfield,\\
Southampton SO17 1BJ, U.K.}
\maketitle
\begin{abstract}
We investigate the nonperturbative behaviour of the quark propagator
in axial gauge using a truncation of the Dyson-Schwinger equations
which respects the Ward-Takahashi identity and multiplicative
renormalisability. We demonstrate that above a critical coupling
$\alpha_c$, which depends on the form of the gluon propagator,
one can obtain massive solutions for both explicit and dynamically
generated quark masses. The stability of these is discussed in the
context of the Cornwall-Jackiw-Tomboulis effective action. 
\end{abstract}

\section{Introduction}

\begin{figure}[htb]
\begin{center}
\leavevmode
\hbox{
\epsfxsize=12cm
\epsfbox{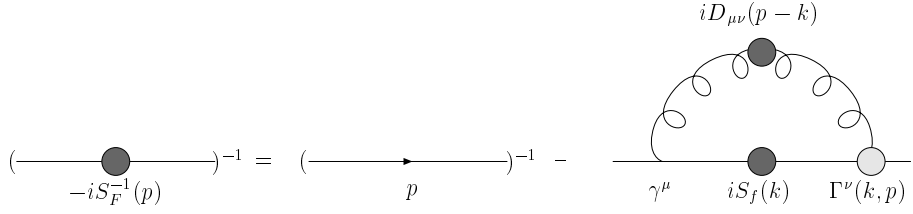}}
\end{center}
\parbox{13cm}{\caption{Diagrammatic representation of the 
  Dyson-Schwinger equation for the inverse quark propagator 
  $S^{-1}(p)$.
\label{DSE-fig}}}
\end{figure}

The Dyson-Schwinger equations (DSEs), in effect the functional
Euler-Lagrange equations of quantum field theory, provide a
natural framework for investigating nonperturbative Green functions.
Whilst this infinite tower of equations encodes all information
relating Green functions of different order, their use was restricted
until quite recently because of the often brutal approximations
which were necessary to render them soluble. A great deal of progress
has been made subsequently, much of which is reviewed in
Ref.~\cite{DSEreview}, which comprehensively covers QED, QCD
and applications to hadronic physics. We will be concerned
with the DSE for the inverse quark propagator, shown diagrammatically 
in  Fig.\ref{DSE-fig}, which relates the fully dressed quark and 
gluon propagators to the complete quark-gluon vertex. Early studies 
replaced the gluon propagator and quark-gluon vertex
by their perturbative expressions, violating the the Ward-Takahashi
identity (WTI) connecting the quark propagator to the vertex and 
destroying the universality of the QCD coupling constant. More 
recently, Bashir and Pennington, extending previous work by Curtis 
and Pennington, have shown how imposing the constraint that the
critical coupling marking the onset of chiral symmetry breaking
be gauge covariant places strong restrictions on
the transverse component of the quark-gluon 
vertex~\cite{ab:mrp,dcc:mrp}. The longitudinal
part can of course be determined from the WTI~\cite{ball:chiu}. 
Whilst studies of DSEs in such a scheme appear to be at an early stage
due to the complex nature of the expressions involved, the weaker
condition of requiring that multiplicative renormalisability be
respected, which also puts strong constraints on the transverse
vertex, has been used widely and with considerable success; 
particularly in massive quenched strongly-coupled 
${\rm QED}_4$~\cite{fth:agw:qed4}.

Lattice simulations hold out the possibility of a first principles
determination of nonperturbative propagators~\cite{nico}. 
However, their use has been hampered because with 
current technology they are unreliable
below $\sim 1$ GeV, where the most interesting and novel
behaviour is expected
to lie. Additionally, gauge fixing on the lattice reduces the number
of gauge configurations available for analysing whilst at the same
time introducing ambiguities due to the presence of Gribov copies.
The latter could be avoided by implementing an axial gauge fixing
procedure on the lattice, rather than the more usual Landau
gauge, but unfortunately there are then further complications
associated with imposing periodic boundary conditions.
It would seem therefore that at the present time DSEs are the most
reliable tool for studying the infrared behaviour of propagators
in the continuum limit.

In what follows we consider the quark propagator $S(p)$, 
given in terms of functions $F(p)$ and $M(p)$ by 
\begin{equation}
S^{-1}(p) = \frac{\gamma.p-M(p)}{F(p)}
\label{propagator}
\end{equation}
and with the form for
the quark-gluon vertex specified by Curtis and 
Pennington~\cite{dcc:mrp}. In
axial gauge the propagator may have additional structure of the
the type $\gamma.n(\gamma.p\,G + H)$, $n$ being the four-vector
appearing in the axial gauge fixing condition $n_\mu A^\mu=0$. The
addition of such terms, although potentially important, 
results in a large increase in the complexity
of the problem, and represents a direction for future research.
We choose also to work with $p.n=0$ as a further condition on
the specification of the gauge, again for practical reasons
to make the equations more tractable. The general gauge dependence
of the quark propagator is extremely interesting, but lies beyond
the scope of present DSE studies except in the case of 
${\rm QED}_4$ in covariant gauges, where details are now beginning
to emerge~\cite{adnan,fth:agw:cdr}

\section{The Quark DSE}

The quark DSE for $S(p)$ at spacelike momenta, $-p^2>0$,
may be written formally as
\begin{equation}
1=\gamma.p\,S(p) -\frac{\alpha C_f}{4\pi} \int D_{\mu\nu}(k-p) S(k)
	\Gamma^\nu(k,p)S(p) \, d^4k,
\label{quark_DSE}
\end{equation}
where $\alpha$ is the strong coupling, $C_f$ is the quark
Casimir, $D_{\mu\nu}(k-p)$ is the full gluon propagator and 
$\Gamma^\nu(k,p)$ is the quark-gluon vertex. The ansatz which we 
shall use for the latter
function is
\begin{equation}
\Gamma^\nu(k,p)=\Gamma^\nu_L(k,p)+\Gamma^\nu_T(k,p),
\end{equation}
with
\begin{eqnarray}
\Gamma^\nu_L(k,p)&=&
{\gamma^\nu \over 2} \left({1\over {F(k)}}+{1\over {F(p)}} \right)
   + {1\over 2} \frac{(k+p)^\nu(\gamma.k+\gamma.p)}{k^2-p^2}
		\left({1\over {F(k)}}-{1\over {F(p)}} \right)
	\nonumber \\
   &-& \left({{M(k)}\over{F(k)}} -{{M(p)}\over{F(p)}} \right) 
		\frac{(k+p)^\nu}{k^2-p^2}	
\end{eqnarray}
and
\begin{eqnarray}
\Gamma^\nu_T(k,p)&=&
{1\over 2}\left({1\over {F(k)}}-{1\over {F(p)}} \right)
	\frac{\gamma^\nu(k^2-p^2)-(k+p)^\nu(\gamma.k-\gamma.p)}{D(k,p)},
\label{qg_vertex}
\end{eqnarray}
in which $D(k,p)$ is a regular function of $k$ and $p$ which behaves
as $k^2$ for $k^2 \gg p^2$. The transverse contribution $\Gamma_T$
cannot be specified uniquely. However, its form may be strongly 
constrained by demanding that the quark propagator respects the
requirements of multiplicative renormalisability, as described by
Curtis and Pennington. We have described previously the
insensitivity of solutions of the quark DSE to the precise form
of $D(k,p)$~\cite{us} and will adopt here
\begin{equation}
D(k,p) = \frac{(k^2-p^2)^2+\left[M^2(k)+M^2(p)\right]^2}
	{k^2+p^2},
\label{arbitrary_D}
\end{equation}
which guarantees that multiplicative renormalisability is respected 
to all orders in leading- and next-to-leading logarithms.
This is the form suggested by Curtis and Pennington in their
important study of the structure of the transverse vertex
and is the most widely used in recent DSE work. Upon 
substituting expressions (\ref{qg_vertex}) for the vertex into the 
quark DSE, it is clear that in the absence of any mass terms, i.e. 
with $M(p)=0$, there is a single linear integral equation for the 
wavefunction $F(p)$ which looks generically like
\begin{equation}
1 = F(p) - \beta \int^\Lambda 
   \frac{\Delta_1(k,p) F(p) + \Delta_2(k,p) F(k)}{k^2} 
   Z\left((k-p)^2\right) d^4k,
\label{fredholm}
\end{equation}
where the shorthand $4\pi\beta=\alpha C_f$ has been used
and the kernels $\Delta_1(k,p)$ and $\Delta_2(k,p)$ are functions
to be specified. $Z(k^2)$ represents the nonperturbative aspect
of the gluon propagator, as will be described in more detail
in the next section. The form of equation (\ref{fredholm})
follows from noting that 
the integral on the right-hand side of the DSE is proportional 
to $F(p) F(k)$, whilst the vertex contains terms which are 
inversely proportional to $F(p)$ or $F(k)$ separately. $\Lambda$
is an ultraviolet regulator in the guise of a sharp momentum cutoff.

In earlier work we elected to evaluate the angular integrals
analytically by means of a shift in variables to remove the angular
dependence from the gluon propagator and then by assuming that $F(p)$
was sufficiently flat that the approximation $F\left((k-p)^2\right)
\approx F(k^2+p^2)$ could be made~\cite{us}. 
This idea had been used previously in a study of the gluon 
propagator in axial gauge~\cite{schoenmaker}, but its validity 
is highly questionable and we shall show that it
leads to spurious errors in the solution for $F(p)$.

We have indicated elsewhere how unique, finite 
solutions to (\ref{fredholm}) exist provided that
the function $\phi(p)$ defined by
\begin{equation}
\phi(p)=1-\beta\int \frac{\Delta_1(k,p)}{k^2} Z(k^2) \,d^4k
\end{equation}
is nonzero for all $p^2$~\cite{us}. Since $\Delta_1(k,p)$ 
and therefore $\phi(p)$ is
independent of $F(p)$, this implies the existence of a value of $\beta$
for which $\phi$ can be made zero and which results in the
breakdown of solutions for $F(p)$ with $M(p)=0$. If the integral
in $\phi(p)$ is positive, this $\beta$ defines a critical value
of the strong coupling $\alpha$ which is interpreted as
signalling chiral symmetry breaking. When the mass function $M(p)$
is reintroduced, $\phi(p)$ is changed to become
\begin{equation}
\phi'(p) = 1-\beta\int \frac{\Delta_1(k,p)}{k^2+M^2(k)}Z(k^2) \, d^4k,
\end{equation}
which has an integrand suppressed at the origin relative to $\phi(p)$
provided $M(k^2\rightarrow 0) \neq 0$. This lifts the condition
$\phi(p)=0$ and means that solutions for $F$ can be recovered
beyond the critical coupling exhibited above. $M(p)$ may be 
introduced by hand as an explicit quark mass, or it may be generated
by the dynamics of the quark-gluon interactions.

The $F(p)$ appearing in equation~(\ref{fredholm}) is a bare quantity
which has an inherent dependence on the ultraviolet regulator
$\Lambda^2$ and should be written strictly as $F(p^2;\Lambda^2)$. Its
subtractive renormalisation follows from introducing a renormalisation 
constant $F_\mu(\mu^2,\Lambda^2)$ so that 
$\widetilde{F}(p^2;\mu^2) = F_\mu(\mu^2,\Lambda^2) F(p^2;\Lambda^2)$
is the renormalised wavefunction which depends on $p^2$ and a 
renormalisation point $\mu^2$ at which we impose the boundary
condition $\widetilde{F}(\mu^2;\mu^2)=1$, corresponding to the
free propagator. The numerical solution of the resulting integral
equation for $F(p)$ is amenable to a number of methods. The one
used here is to map the integration over $k$ onto a finite logarithmic
mesh, which entails both infrared and ultraviolet cutoffs whose
effect will need to be determined. Following this the integral
is replaced by a discrete sum so that
\begin{equation}
\int K(p,k) f(k) \, dk \approx \sum_{j=1}^N K(p_i,k_j) f(k_j) \,
	\omega_j
\label{gauss_quad}
\end{equation}
where the weights $\omega_j$ are chosen according to a Gaussian
quadrature rule and $N$ is the number of grid points used. 

At each value of $(p,k)$ it is necessary to have also a numerical 
evaluation of the angular integrals since the gluon contains the 
dependence $Z\left((k-p)^2\right)=Z(k^2+p^2-2pk\cos\theta)$. With
these steps, the solution of (\ref{fredholm}) becomes equivalent to
determining the elements $f(p_i)$ which satisfy $N$ simultaneous
linear equations of the general type
\begin{equation}
f(p_i) + \sum_{j=1}^NK(p_i,k_j)f(k_j)\,\omega_j=g(p_i).
\end{equation}
The accuracy of the discretised approximation can be found
retrospectively by iterating the equation for $F(p)$ with increasing
$N$ to ensure that the error is small.

%%%%%%%%%%%%%%%%%%%%%%%%%%%%%%%%%%%%%%%%%%%%%%%%%%%%%%%%%%%%%%%%%%%%
\section{Chiral Solutions}

The kernels $\Delta_1(p,k)$ and $\Delta_2(p,k)$ do not depend on
either $F(p)$ or $Z(k^2)$ and are given in the appendix. 
An expansion of $\Delta_1$ and $\Delta_2$ for $k^2\gg p^2$ 
reveals that the integrand involving $\Delta_1$
decreases as a power $(k^2)^{-\delta}Z(k^2)$ with $\delta=3/2$ and the
integrals involving it are finite. On the other hand $\Delta_2$
behaves as $(k^2)^{0}$ for $k^2\gg p^2$ and the finiteness of
its contribution depends on the asymptotic behaviour of $F(k)$
and $Z(k^2)$. We must specify the precise form of $Z(k^2)$ to
be used to determine the critical coupling and numerical solutions
for $F(p)$. An additional point worth noting is that in a scheme where
MR is not enforced, the divergence in the equations is present in the
$\Delta_1$ rather than the $\Delta_2$ terms. 
The coefficient of the divergence is a function of $p^2$ and
cannot be removed by a straightforward subtraction.

In one of the first serious studies of the gluon DSE, Baker, Ball
and Zachariasen (BBZ) examined the consequences of a gluon propagator
which had the same tensor structure as the perturbative one
combined with the scalar function $Z(k^2)$ representing the
nonperturbative dynamics so that~\cite{BBZ:gluon}
\begin{equation}
D_{\mu\nu}(k^2) = -\frac{Z(k^2)}{k^2} \left( g_{\mu\nu}
  - \frac{k_\mu n_\nu + k_\nu n_\mu}{k.n} 
  + n^2 \frac{k_\mu k_\nu}{(k.n)^2} \right).
\label{gluon_prop}
\end{equation}
More generally in axial gauge the propagator could contain the 
additional contribution
\begin{equation}
D'_{\mu\nu}(k^2) = \rho_2(k^2) \left( g_{\mu\nu} 
  -\frac{n_\mu n_\nu}{n^2} \right),
\end{equation}
where $\rho_2(k^2)$ is a further function to be determined. The value
of taking (\ref{gluon_prop}) alone is that the terms in the gluon
DSE involving the full four-gluon vertex are projected out,
resulting in a considerable simplification. Of course, the main
attraction of axial gauge is the absence of the ghost fields, which
removes one of the diagrams involved in the gluon equation and
means that the Ward-Slavnov-Taylor identity for the quark-gluon
vertex can be replaced by the Ward-Takahashi identity, enabling us
to use the expression for the vertex in (\ref{qg_vertex}), which
was originally proposed for QED, without further approximation.
The solutions found by BBZ, which incorporated only the longitudinal
contribution to the quark-gluon vertex, 
exhibited a $Z(k^2)$ which diverged
strongly as $(k^2)^{-1}$ in the infrared and which had the
asymptotic limit $(\ln k^2)^{-11/16}$, which disagrees with 
renormalisation group improved one-loop perturbation theory.
However, in attempts to describe phenomenology by the exchange of 
one or more soft gluons, infrared divergences arise unless the
gluons are softer than $(k^2)^{-1}$ as $k^2\rightarrow 0$. On
the other hand, the axial gauge propagator cannot be nonsingular
at $k^2=0$. The remaining possibility is for a cut solution
which behaves as $(k^2)^{-\zeta}$ with $0<\zeta<1$.
The nonlinearity of the gluon DSE
suggests that there may exist alternative solutions even within
the same truncation schemes.

A very successful phenomenology of diffractive scattering, elastic
scattering, the gluon structure function and total cross-sections
results from the idea, first suggested by Landshoff and Nachtmann
\cite{lands:nacht}, of a process-independent softening of the
gluon propagator in the infrared domain~\cite{phenom}.
Models of diffractive scattering, such as two-gluon exchange,
suffer from infrared divergences if the propagator is not softer
than $1/k^2$ as $k^2\rightarrow 0$. In particular, it is not possible
to obtain the experimentally observed slope of the hadronic scattering
cross-sections in the $t$-channel with an extremely infrared-divergent
gluon. Whilst a strongly infrared-singular propagator might well be
relevant to the question of the nature of confinement, a softer 
behaviour is far more phenomenologically efficacious.

With this is mind, we wish to examine the consequences of a gluon 
propagator which is less singular than the one suggested by BBZ
and which may therefore be phenomenologically relevant. Several 
alternatives are considered. As indicated above, it is believed that
the gluonic wavefunction cannot be singular at the origin and
the perturbative (quenched) propagator with $Z_a(k^2)=1$ thus provides
one limiting case. At the other extreme, one could have a propagator 
with no singularity at the origin, i.e. with $Z(k^2)=k^2$ as 
$k^2\rightarrow 0$. Of course, the gluon cannot behave like this in 
the ultraviolet regime and it is necessary to introduce a scale
$\mu^2$ which marks the transition from the nonperturbative to the 
perturbative region. Perturbatively, $Z(k^2)$ runs as 
$[\ln k^2]^{-1}$ asymptotically. In order to reproduce this behaviour
and to have $Z(k^2=\mu^2)$ equal for all choices of the propagator,
we take as our second ansatz
\begin{equation}
Z_b(x^2) = 
\cases{x^2,	             &$x^2\le 1$ \cr
[\ln (e-1+x^2)]^{-1},        &$x^2>1$ \cr}
\label{Zb}
\end{equation}
with $x^2\mu^2=k^2$.

$Z_a$ and $Z_b$ represent two extreme alternatives and an intermediate
between them is required. Ideally this should share the correct UV
behaviour of $Z_b$ and lie between $Z_a$ and $Z_b$ for $k^2<\mu^2$.
One possibility, which does have these properties, has
phenomenologically desirable infrared behaviour and which can be 
applied successfully to reproduce the experimental data for $p-p$
scattering is that proposed by Cudell and Ross~\cite{jr:doug}:
\begin{equation}
Z_c(x^2) = \frac{x^2}{0.88(x^2)^{0.22} - 0.59(x^2)^{0.86}
  + 0.95 x^2 \ln(2.1 x^2+4.1)}.
\label{CR_gluon}
\end{equation}
This was derived originally from a study of the nonlinearity of the
gluon DSE, which admits the existence of more than one solution. 
That particular analysis is flawed because of a sign 
error~\cite{buttner:mrp}, but the
phenomenologically attractive properties of equation (\ref{CR_gluon})
plus its position between $Z_a$ and $Z_b$ in the infrared domain
make it an ideal ansatz for present puurposes.

\begin{figure}[htb]
\begin{center}
\leavevmode
\hbox{
\epsfxsize=12cm
\epsfbox{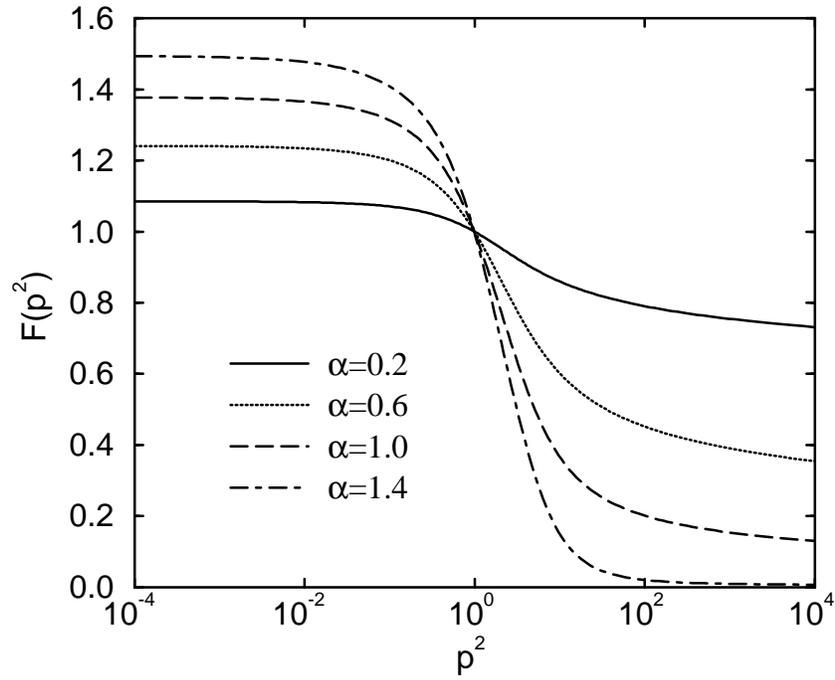}}
\end{center}
\parbox{13cm}{\caption{Solutions to the massless quark DSE with
 the Curtis-Pennington vertex and Cudell-Ross gluon propagator
 at the couplings $\alpha=0.2, 0.6, 1.0, 1.4$. The critical
 value in this scheme was found to be $\alpha_c\approx 1.45$.
\label{CR_massless_fig}}}
\end{figure}

The broad shape of the curves obtained for $F(p)$ is not affected
by $Z(k^2)$. They are illustrated in Fig.~\ref{CR_massless_fig}
with the choice $Z(k^2)=Z_c(k^2)$, for $\alpha=0.2,0.6,1.0,1.4$.
These show a plateau for $p^2 \leq \mu^2$, whose magnitude increases
with $\alpha$, passing through the fixed point at $F(\mu^2)=1$
and thereafter decreasing logarithmically as $(\ln p^2)^{-\delta}$
with $\delta\approx 0.2$. The magnitude of $F(p)$ at large $p^2$
is suppressed as $\alpha$ is increased. In fact the value of
$\alpha$ when $F(p)$ crosses the axis is coincident with the
loss of solutions and corresponds to the critical coupling.
In this case the critical value $\alpha_c$ was found to be
$\alpha_c\approx 1.45$. For the alternatives for $Z(k^2)$,
$Z_a$ and $Z_b$ with the infrared limits $Z_a\rightarrow 1$ and
$Z_b\rightarrow k^2$, we found $\alpha_c=0.9$ and $\alpha_c=1.9$
respectively. Interestingly the former of these is close to
the $\alpha_c\approx 0.93$ which has been found in Landau gauge
investigations of the quark propagator with a 
quenched gluon. When previously we used
the approximation $F((p-k)^2) \approx F(p^2+k^2)$ to derive
the integral equation for $F$, we found oscillations in the region
$p^2\sim\mu^2$, whose magnitude grew with the coupling $\alpha$
until they dominated the solutions at $\alpha\sim\alpha_c$.
They are completely absent from the present set of solutions
and are thus clearly an artifact of the approximation. This is not
too great a surprise, for whilst the derivative of $F(p)$ is
indeed small for $p^2\ll\mu^2$ and $p^2\gg\mu^2, F(p)$ changes
rapidly near $p^2=\mu^2$.

The value of $\alpha_c$ is dominated by the infrared behaviour
of the gluon and reduces with an increasingly singular gluon, 
which is an interesting effect in its own right. It suggests that
the more rapidly the gluonic wavefunction increases at large
distances, the lower the coupling at which chiral symmetry
is dynamically broken. A determination as to whether the behaviour 
of the gluon is the 
major factor contributing to chiral symmetry breaking would require
a more detailed analysis in which the quark and gluon DSEs were
solved simultaneously. Conversely, it is quite plausible that
an even softer gluon would not support chiral symmetry breaking
at all. The test of whether the breakdown of chiral symmetry
is the correct interpretation of $\alpha_c$ will follow from 
examining the more general case of a massive propagator.

%%%%%%%%%%%%%%%%%%%%%%%%%%%%%%%%%%%%%%%%%%%%%%%%%%%%%%%%%%%%%%%
\section{The Massive DSE}

Apart from the obvious complication of adding mass terms to the
quark DSE, a feature which has been neglected frequently is the
correct renormalisation prescription in the presence of a chiral
symmetry breaking mass. This is nontrivial and of great importance
to a fully self-consistent analysis of the solutions for $F(p)$
and $M(p)$. It is worth presenting the details here and we follow
the notation used in Ref.\cite{fth:agw:cdr}. In what follows,
tildes denote
renormalised quantities and primes denote regularised ones which
depend on the UV regulator $\Lambda$.

The starting point is the renormalised inverse quark propagator
$\widetilde{S}^{-1}(p)$ which can be written in terms of the 
renormalised current mass $m(\mu)$ and renormalised self-energy 
function $\widetilde\Sigma(p;\mu)$ as
\begin{equation}
\widetilde{S}^{-1}(p) = \gamma.p-m(\mu)-\widetilde\Sigma(p;\mu) 
  = \left[\gamma.p-m_0(\Lambda)\right] Z_2(\mu,\Lambda)
  - \Sigma'(p;\mu,\Lambda)
\label{prop_def}
\end{equation}
where all the dependencies have been shown explicitly,
$Z_2(\mu,\Lambda)$ is the fermion renormalisation constant and 
$m_0(\Lambda)$ is the bare mass, which is zero in the massless
limit of QCD. The self-energy can be decomposed into Dirac
and scalar components as $\widetilde\Sigma(p;\mu,\Lambda)
=\gamma.p\,\widetilde\Sigma_1(p;\mu,\Lambda)
+\widetilde\Sigma_2(p;\mu,\Lambda)$ with a similar expression
for $\Sigma'$. At the renormalisation point the boundary condition
\begin{equation}
\left.\widetilde{S}^{-1}(p)\right|_{p^2=\mu^2}
  = \gamma.p -m(\mu)
\label{boundary}
\end{equation}
is imposed. Upon inserting the decomposed self-energy into
(\ref{prop_def}) and using (\ref{boundary}), a comparison
of the scalar terms and those involving $\gamma$ matrices 
reveals the relations between the renormalisation constant
$Z_2, \Sigma_1, \Sigma_2$ and the bare and renormalised masses to be
\begin{eqnarray}
Z_2(\mu,\Lambda)= 1+\Sigma_1'(\mu;\mu,\Lambda),  \nonumber \\
Z_2(\mu,\Lambda)m_0(\Lambda)=m(\mu)-\Sigma_2'(\mu;\mu,\Lambda).
\label{Z2_identities}
\end{eqnarray}
One can see from the second of these
that in general for finite $\Lambda, m(\mu)=0$ does 
not necessarily imply that $Z_2(\mu,\Lambda)m_0(\Lambda)=0$ --
this only happens when $m(\mu)=\Sigma_2'(\mu;\mu,\Lambda)$.
As the cutoff is removed however it is necessary that
$\lim_{\Lambda\rightarrow\infty}Z_2 m_0=0$ to ensure conservation
of the axial current $j_5^\mu$ whose divergence is
$\partial_\mu j_5^\mu=\partial_\mu(\bar\psi\gamma^\mu\gamma_5\psi)$. 
In practice we are forced to adopt a finite
cutoff in numerical simulations, but one can check the dependence
of $Z_2m_0$ on $\Lambda$ to verify that it does tend to zero in the
limit.

Using (\ref{prop_def}) together with the relations between bare and
renormalised quantities in (\ref{Z2_identities}) it is 
straightforward to show that the relation between the renormalised
and regularised self-energies is
\begin{equation}
\widetilde\Sigma(p;\mu) = \Sigma'(p;\mu,\Lambda)
   -\Sigma'(\mu;\mu,\Lambda).
\label{sigma_renorm}
\end{equation}
Writing the quark DSE in terms of renormalised quantities
and recalling that in axial gauge $Z_1=Z_2$ we can see that the
regularised self-energy is given by
\begin{equation}
\Sigma'(p;\mu,\Lambda)=iZ_2(\mu,\Lambda)\frac{g^2 C_f}{(2\pi)^4}
  \int^\Lambda d^4k \, \gamma^\rho \widetilde{S}(k) 
  \widetilde\Gamma^\sigma(k,p) \widetilde{D}_{\rho\sigma}(k-p).
\label{sigma_regul}
\end{equation}
Taking the trace of both sides, or the trace after multiplying 
throughout by $\gamma.p$ yields a pair of coupled nonlinear 
equations for the two components of $\Sigma'$. We wish to find
the propagator in the form specified by (\ref{propagator}),
in terms of which the renormalisation condition for $M(p)$ is
\begin{equation}
\frac{M(p;\mu)}{\widetilde{F}(p;\mu)} = m(\mu)
+ \frac{M(p;\mu,\Lambda)}{F'(p;\mu,\Lambda)}
- \frac{M(\mu;\mu,\Lambda)}{F'(\mu;\mu,\Lambda)}.
\label{M_renorm}
\end{equation}
%%
  
%%%%%%%%%%%%%%%%%%%%%%%%%%%%%%%%%%%%%%%%%%%%%%%%%%%%%%%%%%%%%%%
\subsection{Explicit Quark Mass}

Explicit chiral symmetry breaking (ECSB) can be induced by
introducing a non-zero renormalised quark current mass $m(\mu)$.
Performing the traces in the DSE produces an equation for $F(p)$
of the form
\begin{eqnarray}
1-\beta\int\frac{\Delta'_1(\mu,k)}{k^2+M^2(k)} \, dk
&=& F(p) 
- \beta F(p)\int\frac{\Delta'_1(p,k)}{k^2+M^2(k)} \, dk
\nonumber \\
&-& \beta\int\frac{\Delta'_2(p,k)-\Delta'_2(\mu,k)}
	{k^2+M^2(k)} F(k) \,dk,
\label{F_nonlin}
\end{eqnarray}
which remains linear in $F$, and a nonlinear equation for $M$
of the form
\begin{eqnarray}
M(p) = m(\mu)F(p) 
&+& \beta Z_2(\mu,\Lambda) F(p) \int d^4k\,
  \frac{\chi_1(p,k) M(p)+\chi_2(p,k) M(k)}{k^2+M^2(k)} \nonumber \\
&-& \beta Z_2(\mu,\Lambda)F(p) \int d^4k\,
  \frac{\chi_1(\mu,k)m(\mu)+\chi_2(\mu,k) M(k)}{k^2+M^2(k)}.
	\nonumber \\
\label{M_nonlin}
\end{eqnarray}
The kernels $\Delta'_1, \Delta'_2, \chi_1$ and $\chi_2$ have been
relegated once more to the appendix and we will refer to the 
denominator $k^2+M^2(k)$ as $D_1(k,M)$ or simply as $D_1$.
A complete determination of the quark propagator now entails the
simultaneous solution of the two integral equations (\ref{F_nonlin})
and (\ref{M_nonlin}). The one for $F$ is simpler as it is still 
linear and can be solved, given a particular $M$, using the method
outlined for the massless case. Equation~(\ref{M_nonlin}) for $M$, 
being nonlinear,
is more difficult and the approach taken was to introduce
a generalisation in the form of a parameter $\lambda$ into the
denominator $D_1$ which now becomes $D_1=k^2+\lambda M^2(k)$. The
advantage of this is that $\lambda=0$ defines a linearised
equation which is easy to solve and we can then increment
$\lambda$ from 0 to 1, solving iteratively at each stage, to
obtain the required solutions. The equations for $F$ and $M$
are solved alternately so that a given $F^{(i)}$
is used to obtain an approximation $M^{(i)}$ which is put back
into the equation for $F$ to find $F^{(i+1)}$ etc. until convergence
of both functions is achieved. The criterion used was that the 
error function defined by
\begin{equation}
\xi_j = \left| \frac{F_j^{(i)} - F_j^{(i+1)}} 
 {F_j^{(i)} + F_j^{(i+1)}} \right|
\end{equation}
satisfied $\xi_j < 10^{-6}$ for all points $p_j$ at which
$F$ was determined, with an equivalent condition for $M$.

\begin{figure}[htb]
\begin{center}
\leavevmode
\hbox{
\epsfxsize=12cm
\epsfbox{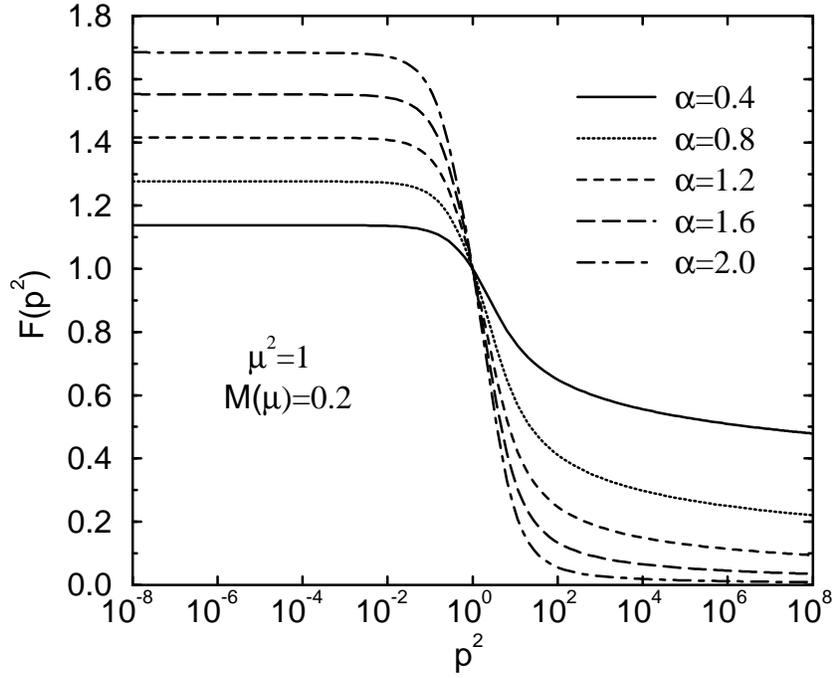}}
\end{center}
\parbox{13cm}{\caption{The wavefunction $F$ with an explicit
 quark mass, $m(\mu)=0.2$ for sub- and super-critical values
 of the coupling $\alpha$.
\label{E-F_fig}}}
\end{figure}

\begin{figure}[htb]
\begin{center}
\leavevmode
\hbox{
\epsfxsize=12cm
\epsfbox{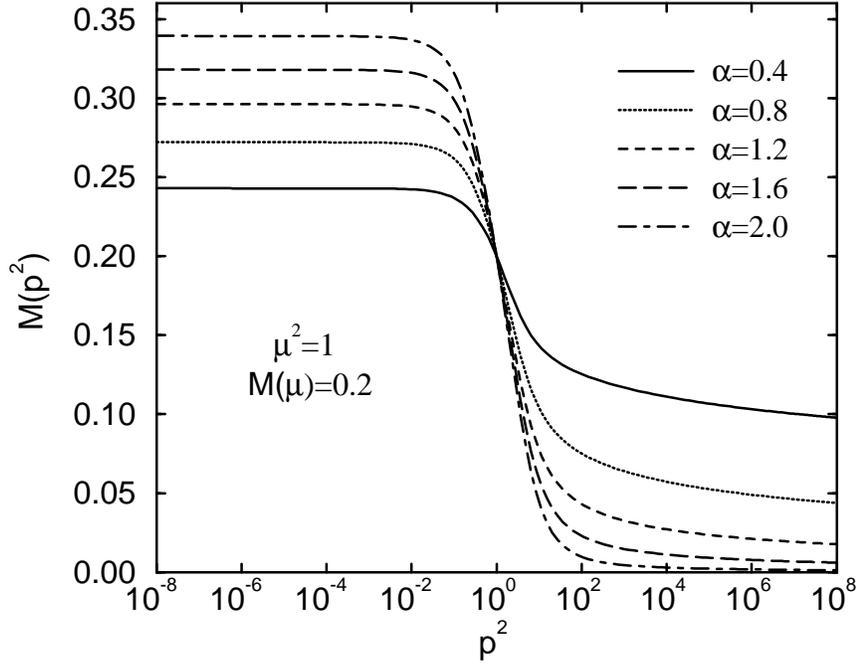}}
\end{center}
\parbox{13cm}{\caption{Mass function $M(p)$ with an explicit 
 current quark mass, $m(\mu)=0.2$.
\label{E-mass_fig}}}
\end{figure}

As an illustration of the functions obtained, the equations
were solved for $\mu^2=1$ with $m(\mu)=0.2$ using the
Cudell-Ross form for the gluon propagator. The wavefunction
$F$ and mass function $M$ are shown in Figs.\ref{E-F_fig} and
\ref{E-mass_fig} respectively for $\alpha=0.4, 0.8, 1.2, 1.6, 2.0$.
The cutoffs used were $\Lambda_{\rm IR}^2=10^{-6}$ and
$\Lambda_{\rm UV}^2=10^6$ (the infrared cutoff was not
expected to be important because all the integrands are strongly
suppressed near $k^2=0$ and indeed there was no significant
variation in any of the quantities calculated for
$\Lambda_{\rm IR}^2=10^{-6}, 10^{-8}, 10^{-10}$ or $10^{-12}$).

The first point of note is that as advertised, incorporating
$M(p)$ does allow us to recover solutions for $F(p)$ for
$\alpha>\alpha_c\sim 1.45$. The magnitude of $M(p)$ as 
$p^2\rightarrow 0$ increases as $\alpha$ is increased, whilst the
tail decreases with increasing $\alpha$ and becomes close to
zero, although it appears never to cross the axis. Second,
for a given $\alpha<\alpha_c$, $F(p)$ is both infrared suppressed
and ultraviolet enhanced relative to the corresponding massless
solution. Third, the large-$p^2$ variation of 
$F(p)$ with respect to $\alpha$ is reduced relative to the
$M=0$ solutions.

\begin{figure}[htb]
\begin{center}
\leavevmode
\hbox{
\epsfxsize=12cm
\epsfbox{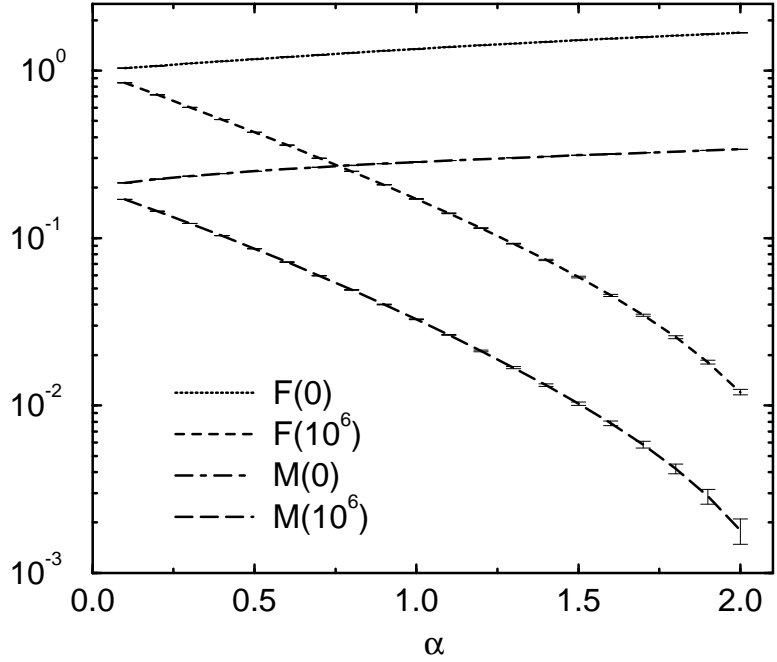}}
\end{center}
\parbox{13cm}{\caption{Estimated variation of $F(0), F(10^6), 
 M(0)$ and $M(10^6)$ due to a finite UV cutoff, as a function
 of the coupling. Error bars represent $\pm 1$ standard deviation.
\label{cutoff-dep_fig}}}
\end{figure}

In order to investigate the dependence of the solutions on the
UV cutoff, the equations were re-solved with the same values
of $\mu^2$ and $m(\mu)$ for each of $\log_{10}\Lambda^2
=6.0, 6.6, 7.2... 12.0$ and the values of $M(p^2\rightarrow 0),
M(p^2=10^6), F(p^2\rightarrow 0)$ and $F(p^2=10^6)$
established. These were used to find mean values and deviations
which are shown in Fig.\ref{cutoff-dep_fig} for $\alpha$ ranging
from 0 to 2.0, where the error bars represent $\pm 1$ standard
deviation. The mean value of $M(10^6)$ is the only one to
show appreciable variation, for $\alpha>1.5$. To eliminate
the possibility of a systematic dependence on $\Lambda$, the
correlation $\rho$ between each of the four quantities shown in Fig.
\ref{cutoff-dep_fig} and $\ln\Lambda$ were evaluated. 
$\rho$ can take
values between $\pm 1$ corresponding to perfect positive or
negative correlations. In no case did it exceed ${\cal O}(10^{-3})$,
which confirms that there is little if any $\Lambda$ dependence
in the renormalised solutions.

%%%%%%%%%%%%%%%%%%%%%%%%%%%%%%%%%%%%%%%%%%%%%%%%%%%%%%%%%%%%%%%
\subsection{Dynamical Mass Generation}

The ECSB solutions discussed above where $m(\mu)\neq 0$
are important -- because quarks do after all have a mass -- but
the symmetry breaking has been put in ``by hand''. To explore
the transition between massless and massive phases, we must solve
the equations for $m(\mu)=0$ (the chiral limit) to see if for
$\alpha > \alpha_c$ a mass is generated dynamically. That is 
to say, can quark-gluon and self-interactions generate a mass
from the initially massless theory? For $m(\mu)=0$ the
equation for $M$ becomes
\begin{equation}
M(p)=\beta Z_2(\mu,\Lambda)F(p)\int^\Lambda
\frac{\chi_1(p,k)M(p)+\left[\chi_2(p,k)-\chi_2(\mu,k) \right]M(k)}
     {k^2+M^2(k)},
\label{dyna_M}
\end{equation}
where $\chi_1$ and $\chi_2$ are unchanged, as is the equation for
$F$ apart from terms involving $M(\mu)=m(\mu)=0$ which vanish.
As noted previously, for a finite cutoff, $m(\mu)=0$ does not
imply that the product $Z_2m_0=0$. However, $\chi_2(\mu,k)$ is 
strongly suppressed for large $\mu^2$, leading to only a small 
residual value of $Z_2 m_0$. This provides the basis for the success
of the many studies which neglect the $\mu$-dependent part of the
right-hand
side of (\ref{dyna_M}) and take $\mu^2=\Lambda_{\rm UV}^2$.
They incur the undesirable side effect that $M(p)$ is barely down 
from its infrared magnitude even at $p^2=\Lambda^2$~\cite{dcc:mrp:2},
contrary to what one would expect for a dynamically generated mass
term which should be small in the perturbative regime.

The new equation for $M$ was solved using a Powell hybrid method, 
after discretisation of the integrals, which is a variation
of the Newton-Raphson technique which avoids the evaluation of
derivatives~\cite{powell}. The technique combines the advantages
of a multidimensional root searching method with a minimisation
of sum of squares to improve global convergence.
This is convenient because it can 
be tailored to
exclude the trivial $M=0$ solution. Solving the equations for
$M$ and $F$ iteratively is problematic since for
$\alpha>\alpha_c$, the existence of a solution for $F$ depends 
strongly on
$M$. With a method which incorporates minimisation, we simply
introduce a ``penalty'' to disfavour any intermediate $M$ which
makes the equation for $F$ insoluble.

\begin{figure}[htb]
\begin{center}
\leavevmode
\hbox{
\epsfxsize=12cm
\epsfbox{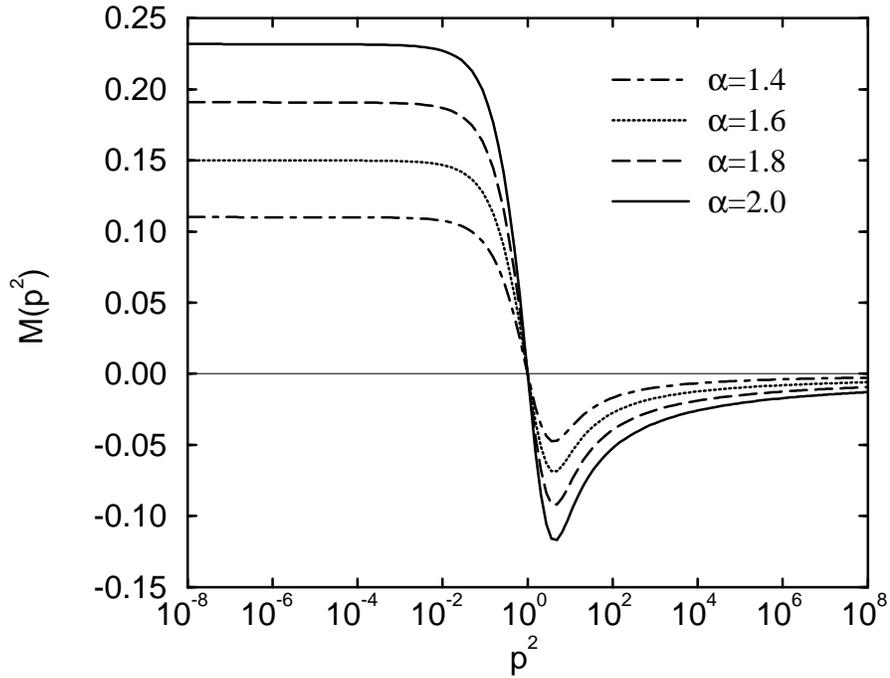}}
\end{center}
\parbox{13cm}{\caption{Dynamically generated quark mass function,
 i.e. with $m(\mu)=0$, for couplings above critical, showing
 strongly damped oscillations.
\label{D-mass_fig}}}
\end{figure}

Once more the renormalisation point $\mu^2=1$ was used at which
$M(p)$ must be zero to satisfy (\ref{M_renorm}). 
Fig.\ref{D-mass_fig} shows the curves for $M$ obtained for
$\alpha$ ranging from 1.4 to 2.0 with $\Lambda^2=10^8$. It was
not possible to find solutions with $M\neq 0$ for $\alpha<1.39$,
which is slightly lower than the critical $\alpha_c=1.45$. For 
$1.39<\alpha<1.45$ there are two possible sets of solutions 
for $F$ and $M$ -- one having $M=0$ and the other $M\neq 0$. In 
Fig.\ref{D-mass_fig} we see that $M$ crosses the axis at
$p^2=\mu^2$, attaining a minimum at $p^2\sim 2$ and then
declining towards zero as $p^2\rightarrow\Lambda^2$. The size
of $M$ in the infrared shows a familiar plateau with $M(0)$
increasing with $\alpha$. This is consistent with preventing
a zero in the function $\phi(p)$ governing the existence of
solutions for $F$. The depth of the minimum increases with
$\alpha$ and $|M_{\rm min}|\approx 0.45\,|M(0)|$.

The corresponding curves for $F$, which are not shown,
are very similar to those in 
Fig.\ref{E-F_fig}, but exhibit an interesting feature, namely
that $F$ is quite insensitive to $\alpha$ in the presence
of a dynamically generated mass. It appears that the dynamical
$M$ holds $F$ close to its massless form immediately below
$\alpha_c$. As $\alpha$ increases, the magnitude of $M$
increases to compensate, keeping $\phi(p)$ slightly above zero.
The oscillations seen in $M$ have also been identified in
strongly coupled massive quenched ${\rm QED}_4$, where they
are much smaller in size. 

\begin{figure}[htb]
\begin{center}
\leavevmode
\hbox{
\epsfxsize=12cm
\epsfbox{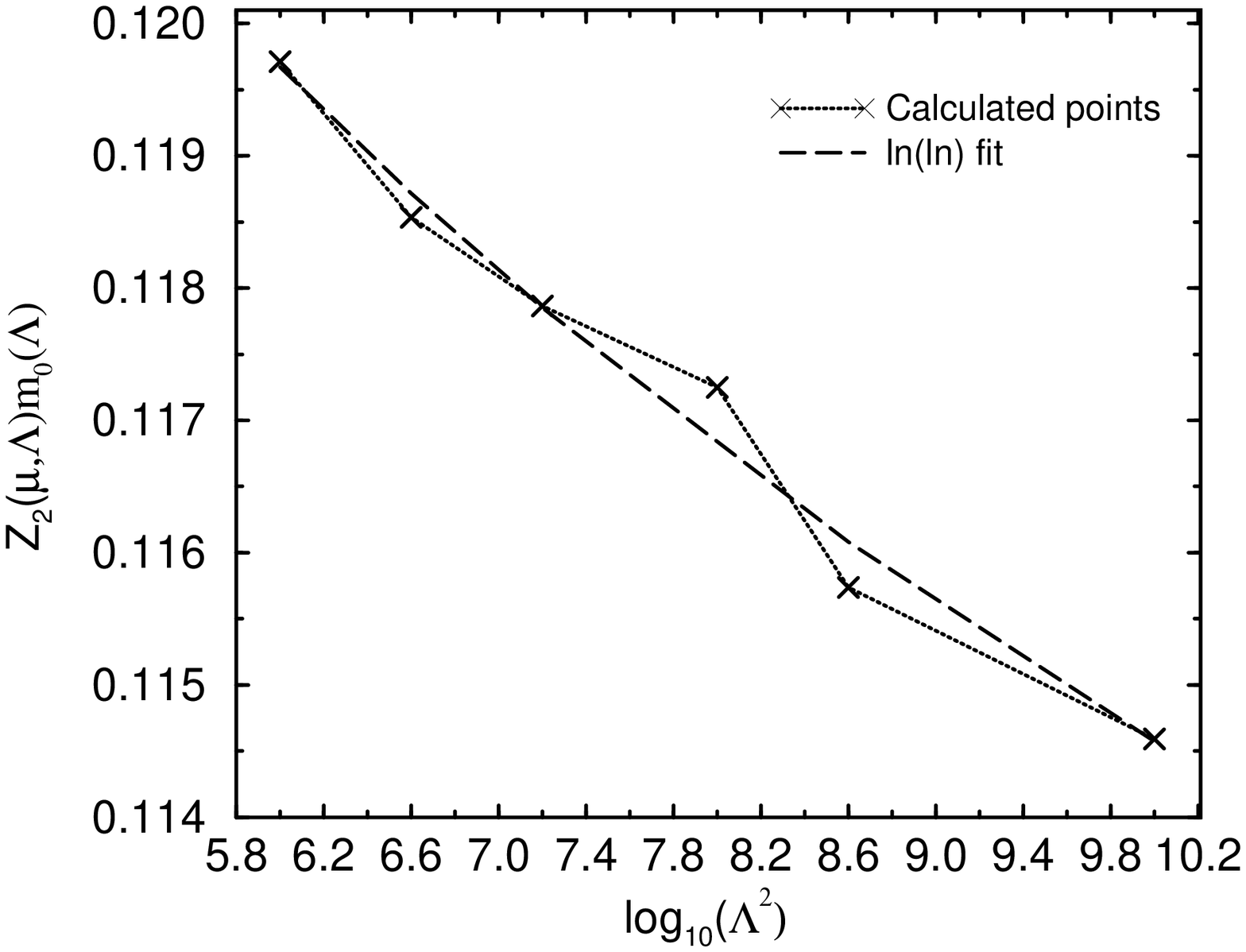}}
\end{center}
\parbox{13cm}{\caption{Variation of 
 $Z_2(\mu,\Lambda)m_0(\Lambda)$ with the UV cutoff $\Lambda$. The
 dashed line is a fit to the predicted $\ln\ln\Lambda$ dependence.
\label{Z2-vari_fig}}}
\end{figure}

The cutoff dependence of the solutions was tested for $\Lambda^2$
ranging between $10^6$ and $10^{12}$ and the renormalised $M$ again
showed no significant change. The product $Z_2m_0$ was computed
at each value of $\alpha$ and $\Lambda^2$ and the variation in 
this is shown in Fig.\ref{Z2-vari_fig}. The slow decrease is
striking and the reason for it lies in the nature of the integrands
involved in calculating $Z_2$. Recalling that 
$Z_2(\mu,\Lambda)=1+\Sigma'(\mu;\mu,\Lambda)$ we have,
after some rearrangement, from (\ref{sigma_regul})
\begin{equation}
Z_2(\mu,\Lambda) = \left\{1+\beta\int^\Lambda
  \frac{\Delta(\mu,k)}{k^2+M^2(k)} \right\}^{-1},
\end{equation}
with $\Delta(p,k)=\Delta_1'(p,k)F(p)+\Delta_2'(p,k)F(k)$.
Expanding the integrand in the case where $k^2\gg\mu^2$ we find
\begin{equation}
\frac{\Delta(p,k)}{k^2+M^2(k)} \sim \frac{2}{k\ln k^2}
 \left[1+F(k) \right],
\end{equation}
where the limits $\theta_1=\theta_3=0$ and 
$2\theta_2=\theta_4=\pi Z(k^2)$ have been used for the angular
integrals (as they appear in the appendix), together with the
leading term in $Z(k^2)\sim \left[k^2\ln k^2\right]^{-1}$.
The dominant part is thus $\Phi(k)\sim\left[k\ln k\right]^{-1}$
and the contribution of this to $Z_2$ is
\begin{equation}
I_\Lambda(\Phi)=\int_a^\Lambda \Phi(p,k) \,dk
=\int_a^\Lambda \frac{dk}{k\ln k}
\end{equation}
with $\Lambda^2\gg a^2\gg \mu^2$, giving 
$I(\Phi)=\ln\ln\Lambda-\ln\ln a$.
So $Z_2 m_0$ does go to zero with increasing $\Lambda$,
but only as a double logarithm -- the very slow decrease shown
in Fig.\ref{Z2-vari_fig} which also shows a line fitting
to $\ln\ln\Lambda$ to demonstrate the close agreement.

%%%%%%%%%%%%%%%%%%%%%%%%%%%%%%%%%%%%%%%%%%%%%%%%%%%%%%%%%%%%%%%
\section{Stability of the Solutions}

The existence of simultaneous chiral symmetry preserving and 
chiral symmetry breaking solutions presents the question as
to which is the physically important stable state. This problem
can be addressed partially through the use of the 
Cornwall-Jackiw-Tomboulis (CJT) effective action 
$\Gamma_{\rm CJT}$~\cite{CJT}. The condition 
$\frac{\delta\Gamma}{\delta S}=0$ locates the stationary points
of the CJT action and is in fact equivalent to the DSE for the
quark propagator. In other words, solutions to the fermion
DSE automatically give $\frac{\delta\Gamma}{\delta S}=0$. At the
stationary point, the CJT effective action can be written in terms
of the propagator $S(p)$ as~\cite{DSEreview}
\begin{equation}
\Gamma_{\rm CJT}[S] = {\rm Tr} \left[\ln S^{-1}\right]
  + {1\over 2}{\rm Tr} \left[1-S_0^{-1}S\right],
\label{CJT-EA}
\end{equation}
which may be evaluated for each of the sets of solutions where
both $M=0$ or $M\neq 0$ are possible. The trace indicates integration
over momenta and summation over all discrete indices. We can 
consider the difference $\Delta\Gamma$ between these, i.e.
\begin{equation}
\Delta\Gamma=\Gamma_{\rm CJT}[M=0] - \Gamma_{\rm CJT}[M\neq 0]
\end{equation}
and a negative value with $\Delta\Gamma < 0$ is interpreted as a 
signal that the occurrence of chiral symmetry breaking solutions
lowers the energy of the vacuum and is preferred physically.
Indeed, when $\Delta\Gamma$ was evaluated for explicit and dynamical
quark masses, it was found that massive solutions are always
favoured. This resolves the ambiguity for $1.39<\alpha<1.45$,
where both a massless quark and one with a dynamically generated
mass were possible. It appears that the latter is favoured.

One could go further and find the derivatives of 
$\Gamma_{\rm CJT}$ to ascertain the nature of the stationary point. 
However the CJT action is unsatisfactory because it is not bounded
below. In particular one can show analytically that in Landau
gauge the solutions to the gap equation with dynamical masses 
are unstable
saddle points of the action although they still have
$\Delta\Gamma<0$ as compared to the chiral solutions.
It is not possible therefore to address conclusively the
matter of the stability through the CJT formalism.

%%%%%%%%%%%%%%%%%%%%%%%%%%%%%%%%%%%%%%%%%%%%%%%%%%%%%%%%%%%%%%
\section{Concluding Remarks}

Axial gauge solutions to the quark DSE have been found, in a
truncation scheme which respects MR, for several possible forms
for the gluon propagator in the infrared domain. These reproduce
the familiar perturbative behaviour at high momentum and exhibit
a critical value for the strong coupling $\alpha_c$, whose value
decreases as the gluon propagator becomes increasingly infrared
singular, independently of its ultraviolet running. Since chiral
symmetry breaking is believed to be connected intimately with
confinement,
this provides further support, should such be required, that
the confinement mechanism is linked closely to the
infrared sector of QCD. 

Above the critical coupling, solutions to the DSE can be recovered
by the introduction of a quark mass, either dynamically or explicitly,
as has been found in numerous covariant gauge studies. The correct
treatment of the renormalisation of the massive case has not always
been handled rigorously in the literature, despite the efforts of
several authors, and yet it is crucially important in obtaining
a self-consistent treatment of the DSE. It is worth reiterating
that in numerical studies which inevitably involve a finite 
infrared regulator, it is not enough to simply set the bare and 
renormalised quark masses simultaneously to zero, although
they must vanish as the cutoff is removed in order to ensure that
the axial current is conserved.

Although we have discussed briefly the question of stability, as
indicated the CJT formalism is not in itself a definitive tool.
There have been a number of attempts to address this
problem through alternative choices for the definition of the
effective action and these represent an important direction for 
future investigation.

Finally, it is known how the ansatz used for the quark-gluon vertex 
can be improved through considerations of gauge invariance. Studies
incorporating a suitable vertex should prove extremely interesting
although they will require the solution of much more complicated 
integral equations. This is particularly so in axial gauge,
and will become especially acute when the extra structure involving
$\gamma.n$ is included in the form for the quark propagator.

%%%%%%%%%%%%%%%%%%%%%%%%%%%%%%%%%%%%%%%%%%%%%%%%%%%%%%%%%%%%%%
\acknowledgements

We would like to thank Jean-Rene Cudell for numerous illuminating
comments and discussions. 

%%%%%%%%%%%%%%%%%%%%%%%%%%%%%%%%%%%%%%%%%%%%%%%%%%%%%%%%%%%%%%%
% Appendix of kernels cropping up in the various cases above. %
%%%%%%%%%%%%%%%%%%%%%%%%%%%%%%%%%%%%%%%%%%%%%%%%%%%%%%%%%%%%%%%     
\begin{appendix}

\section{Integral equation kernels}

Collected together here are the expressions for the various integral
equation kernels which have been encountered earlier. The following
shorthand notation has been adopted for the angular integrals:
\begin{eqnarray}
\theta_1 &=& \int_0^\pi Z(p^2+k^2-2pk\cos\theta) 
	\sin^2\theta \cos\theta \,d\theta, \nonumber \\
\theta_2 &=& \int_0^\pi Z(p^2+k^2-2pk\cos\theta) 
	\sin^2\theta \,d\theta,\nonumber \\
\theta_3 &=& \int_0^\pi Z(p^2+k^2-2pk\cos\theta) 
	\cos\theta \,d\theta, \nonumber \\
\theta_4 &=& \int_0^\pi Z(p^2+k^2-2pk\cos\theta) \,d\theta.
\end{eqnarray}
For the massless case with the Curtis-Pennington vertex one has
\begin{eqnarray}
\Delta_1(p,k) 
&=& - 2p k^2 \theta_3 - 4{k^4\over p}\theta_1 - 2{k^4\over p}\theta_3 
   + 4k^3 \theta_2 + 4k^3 \theta_4                  \nonumber \\
&+& {1\over {D(p,k)}}  
   \left[ - 4p^2 k^3 \theta_2 + 16p k^4 \theta_1 
   - 8{k^5\over p}\theta_1 - 4k^5 \theta_2 \right]  \nonumber \\
&+&  \frac{1}{k^2-p^2} \left[ - 2p^3k^2 \theta_3 + 2p^2k^3 \theta_2 
   + 4pk^4 \theta_1 + 4pk^4 \theta_3 - 2{k^6\over p}\theta_3 
   + 2k^5 \theta_2 \right],                         \nonumber \\
\Delta_2(p,k)
&=& - 2pk^2\theta_3 - 4{k^4 \over p}\theta_1 - 2{k^4\over p}\theta_3 
   + 4k^3\theta_2 + 4k^3\theta_4                    \nonumber \\
&+& {1\over {D(p,k)}}  \left[ 4p^2k^3\theta_2 - 16pk^4\theta_1 
   + 8{k^6\over p}\theta_1 + 4k^5\theta_2 \right]   \nonumber \\
&+&  \frac{1}{k^2-p^2}  \left[ 2p^3k^2\theta_3 - 2p^2k^3\theta_2 
   - 4pk^4\theta_1 - 4pk^4\theta_3 + 2{k^6\over p}\theta_3 
   - 2k^5\theta_2 ) \right].	                    \nonumber \\
\label{CP_massless_kerns}
\end{eqnarray}

In the massive case, the kernels appearing in the equation for $F$
are given by
\begin{eqnarray}
\Delta'_1 
&=& - 2p k^2 \theta_3 - 4{k^4\over p}\theta_1 
  - 2{k^4\over p}\theta_3 + 4k^3 \theta_2 + 4k^3 \theta_4
		\nonumber \\
&+& {1\over D(p,k)}  \left[ - 4p^2 k^3 \theta_2 
  + 16p k^4 \theta_1 - 8{k^5\over p}\theta_1 - 4k^5 \theta_2 \right] 
		\nonumber \\
&+&  \frac{1}{k^2-p^2} \left[ - 2p^3k^2 \theta_3 + 2p^2k^3 \theta_2 
  + 4pk^4 \theta_1 + 4pk^4 \theta_3 - 2{k^6\over p}\theta_3 
  + 2k^5 \theta_2\right] 
		\nonumber \\
&+& \frac{M^2(k)}{k^2-p^2}  \left[ - 4p^2k \theta_4 + 4pk^2 \theta_3 
  - 4{k^4\over p}\theta_3 + 8k^3 \theta_2 + 4k^3 \theta_4 \right]	
		\nonumber \\
\Delta'_2
&=& - 2pk^2\theta_3 - 4{k^4 \over p}\theta_1 - 2{k^4\over p}\theta_3 
  + 4k^3\theta_2 + 4k^3\theta_4
		\nonumber \\
&+& {1\over D(p,k)}  \left[ 4p^2k^3\theta_2 - 16pk^4\theta_1 
  + 8{k^6\over p}\theta_1 + 4k^5\theta_2 \right]	
		\nonumber \\
&+&  \frac{1}{k^2-p^2}  \left[ 2p^3k^2\theta_3 - 2p^2k^3\theta_2 
  - 4pk^4\theta_1 - 4pk^4\theta_3 + 2{k^6\over p}\theta_3 
  - 2k^5\theta_2 )\right] 
		\nonumber \\
&+&  \frac{M(p)M(k)}{k^2-p^2}  \left[ 4p^2k\theta_4 - 4pk^2\theta_3 
  + 4{k^4\over p}\theta_3 - 8k^3\theta_2 - 4k^3\theta_4 \right], 
		\nonumber \\
%\label{F_non_kern}
\end{eqnarray}
and those from the equation for $M$ are
\begin{eqnarray}
\chi_1(p,k) &=& \frac{F(k)}{F(p)}{1\over{k^2-p^2}}
  \left[ - 4p^3k^2\theta_3 + 4p^2k^3\theta_2 + 4p^2k^3\theta_4 
  + 8pk^4\theta_1 \right. \nonumber \\
& & \qquad \mbox{} \left. +\, 4pk^4\theta_3 - 4k^5\theta_2 
  - 4k^5\theta_4 \right],
		\nonumber \\
\chi_2(p,k) &=& \frac{1}{k^2-p^2}  \left[ 4p^3k^2\theta_3 
  - 4p^2k^3\theta_2 - 4p^2k^3\theta_4 - 8pk^4\theta_1 
  - 4pk^4\theta_3 \right. \nonumber \\
& & \qquad \mbox{} \left. + \,4k^5\theta_2 + 4k^5\theta_4 \right] 
	\nonumber \\
&+& {1\over{F(p)}}\frac{F(p)-F(k)}{D(p,k)}  
  \left[ - 4p^2k^3\theta_2 - 8pk^4\theta_1 + 12k^5\theta_2 \right]
	\nonumber \\
&+& {1\over{F(p)}}\frac{F(p)-F(k)}{k^2-p^2}  \left[- 2p^4k\theta_4 
  + 6p^2k^3\theta_2 + 4p^2k^3\theta_4 + 4pk^4\theta_1 
  \right. \nonumber \\
& & \qquad \mbox{} \left. -\, 2k^5\theta_2 - 2k^5\theta_4 \right]
	\nonumber \\
&+& {1\over{F(p)}}\left( F(p)+F(k) \right)
  \left[- 2p^2k\theta_4 + 4pk^2\theta_3+ 4k^3\theta_2 
  - 2k^3\theta_4 \right] 
		\nonumber \\
%\label{M_non_kern}
\end{eqnarray}

\end{appendix}
%%%%%%%%%%%%%%%%%%%%%%%%%%%%%%%%%%%%%%%%%%%%%%%%%%%%%%%%%%%%%%%%

\begin{references}
\bibitem{DSEreview} C.D.~Roberts and A.G.~Williams, Prog. Part. Nucl.
  Phys. {\bf 33}, 477-575 (1994), hep-ph/9403224.
\bibitem{ab:mrp} A.~Bashir and M.R.~Pennington, Phys. Rev. 
  {\bf D50}, 7679-7689 (1994), hep-ph/9407350.
\bibitem{dcc:mrp} D.C.~Curtis and M.R.~Pennington, Phys. Rev.
  {\bf D42}, 4165-4169 (1990).
\bibitem{ball:chiu} J.S.~Ball and T-W.~Chiu, Phys. Rev.
  {\bf D22}, 2542-2549 (1980).
\bibitem{fth:agw:qed4} F.T.~Hawes and A.G.~Williams, Phys. Rev.
  {\bf D51}, 3081-3089 (1995), hep-ph/9410286.
\bibitem{nico} P.~Marenzoni, G.~Martinelli, N.~Stella and M.~Testa,
  Phys. Lett. {\bf B318}, 511-516 (1993).
\bibitem{adnan} A.~Bashir ``Chiral Symmetry Breaking for
  Fundamental Fermions'' in Proceedings of the NATO ASI 
  ``Frontiers in Particle Physics: Carg\`ese 1994'' (1995), hep-ph/9410399.
\bibitem{fth:agw:cdr} F.T.~Hawes, A.G.~Williams and C.D.~Roberts,
  Phys. Rev. {\bf D54}, 5361 (1996), hep-ph/9604402.
\bibitem{us} J-R.~Cudell, A.J.~Gentles and D.A.~Ross, Nucl. Phys. 
  {\bf B440}, 525 (1995), hep-ph/9407220.
\bibitem{schoenmaker} W.J.~Schoenmaker, Nucl. Phys. {\bf B194},
  535 (1982).
\bibitem{BBZ:gluon} J.S.~Ball, M.~Baker and F.~Zachariasen,
  Nucl. Phys. {\bf B186}, 531 (1981); J.S.~Ball, M.~Baker 
  and F.~Zachariasen, Nucl. Phys. {\bf B186}, 560 (1981).
\bibitem{lands:nacht} P.V.~Landshoff and O.~Nachtmann,
  Z. Phys {\bf C35}, 405 (1987); 
\bibitem{phenom} See for example A.~Donnachie and P.V.~Landshoff,
  Phys. Lett. {\bf B185}, 403 (1987); A.~Donnachie and P.V.~Landshoff,
  Nucl. Phys. {\bf B311}, 509 (1989); J-R.~Cudell, A.~Donnachie
  and P.V.~Landshoff, Nucl. Phys. {\bf B322}, 55 (1989).
\bibitem{jr:doug} J-R.~Cudell and D.A.~Ross, Nucl. Phys.
  {\bf B359} 247-261 (1991).
\bibitem{dcc:mrp:2} See for example D.C.~Curtis and M.R.~Pennington, 
  Phys. Rev. {\bf D48}, 4933-4939 (1993).
\bibitem{powell} M.J.D.~Powell, in P.~Rabinowitz (ed.) Numerical
  Methods for Nonlinear Algebraic Equations, Gordon and Breach 
  (1970).
\bibitem{CJT} J.M.~Cornwall, R.~Jackiw and E.~Tomboulis,
  Phys. Rev. {\bf D10}, 2428-2445 (1974).

\bibitem{buttner:mrp} K.~Buttner and M.R.~Pennington, Phys. Rev.
  {\bf D52} 5220-5228 (1995), hep-ph/9506314.
\bibitem{fth:cdr:agw} F.T.~Hawes, C.D.~Roberts and A.G.~Williams,
  Phys. Rev. {\bf D49} 4683-4693 (1994), hep-ph/9309263.
\end{references}
\end{document}